\documentclass[a4paper,twoside]{article}

\usepackage{epsfig}
\usepackage{subcaption}
\usepackage{calc}
\usepackage[smaller,nolist,nohyperlinks]{acronym}
\usepackage{hyperref}
\usepackage{amssymb}
\usepackage{amstext}
\usepackage{amsmath}
\usepackage{amsthm}
\usepackage{multicol}
\usepackage{pslatex}
\usepackage{tikz}
\usepackage{apalike}
\usepackage{algorithm2e}
\usepackage[bottom]{footmisc}
\usepackage{url}
\usepackage{cleveref}
\usepackage{todonotes}
\usepackage{SCITEPRESS}

\usetikzlibrary{patterns}

\newboolean{anonymous} 
\setboolean{anonymous}{false}
\newcommand{\onlyan}[1]{\ifthenelse{\boolean{anonymous}}{#1}{}}
\newcommand{\onlynonan}[1]{\ifthenelse{\boolean{anonymous}}{}{#1}}

\begin{acronym}
	\acro{elk}[ELK]{Eclipse Layout Kernel}
	\acro{klighd}[KLighD]{Kieler Lightweight Diagrams}
	\acro{zui}[ZUI]{Zoomable User Interface}
\end{acronym}

\definecolor{small}{RGB}{0, 156, 34}
\definecolor{medium}{RGB}{247, 120, 0}
\definecolor{large}{RGB}{0, 104, 247}

\definecolor{bottomup}{RGB}{0, 104, 247}
\definecolor{topdown1}{RGB}{255, 59, 20}
\definecolor{topdown2}{RGB}{225, 185, 0}
\definecolor{topdown3}{RGB}{0, 173, 11}

\SetCommentSty{commentFont}

\begin{document}

\title{Top-Down Drawings of Compound Graphs}

\onlynonan{
    \author{\authorname{Maximilian Kasperowski\sup{1}\orcidAuthor{0000-0002-7509-1678} and
	Reinhard~von~Hanxleden\sup{1}\orcidAuthor{0000-0001-5691-1215}}
	\affiliation{\sup{1}Department of Computer Science, Kiel University, Kiel, Germany}
  \email{\{mka,rvh\}@informatik.uni-kiel.de}}
}
\onlyan{
    \author{\authorname{Anon Ymos\sup{1}\orcidAuthor{0000-0000-0000-0000}}
	  \affiliation{\sup{1}Blinded for review}
		\email{anon@nym.os}}
}


\abstract{Bottom-up layout algorithms for compound graphs are suitable for presenting the microscale view of models and are often used in model-driven engineering.
However, they have difficulties at the macroscale where maintaining the overview of large models becomes challenging.
We propose \emph{top-down} layout, which utilizes scale to hide low-level details at high zoom levels.
The entire high-level view can fit into the viewport and remain readable, while the ability to zoom in to see the details is still maintained.
Top-down layout is an abstract high-level layout process that can be used in conjunction with classic layout algorithms to produce visually compelling and readable diagrams of large compound graphs.}

\onecolumn \maketitle \normalsize \setcounter{footnote}{0} \vfill

\section{\uppercase{Introduction}}
\label{sec:introduction}

\begin{figure*}[h!]
	\centering
	\begin{subfigure}{.49\linewidth}
		\centering
		\includegraphics[width=\linewidth]{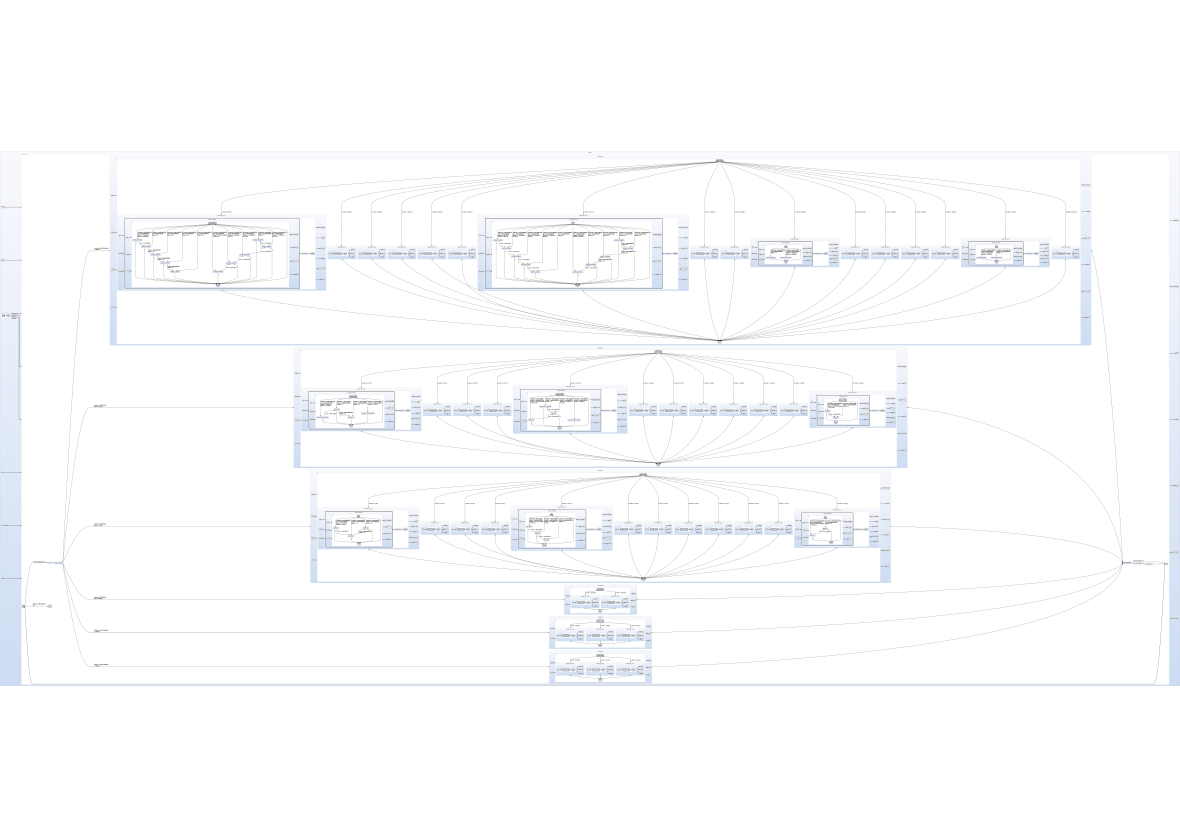}
		\caption{Bottom-up layout.}
		\label{fig:intro-wagon-bottom-up}
	\end{subfigure}
	\hfill
	\begin{subfigure}{.49\linewidth}
		\centering
		\includegraphics[width=\linewidth]{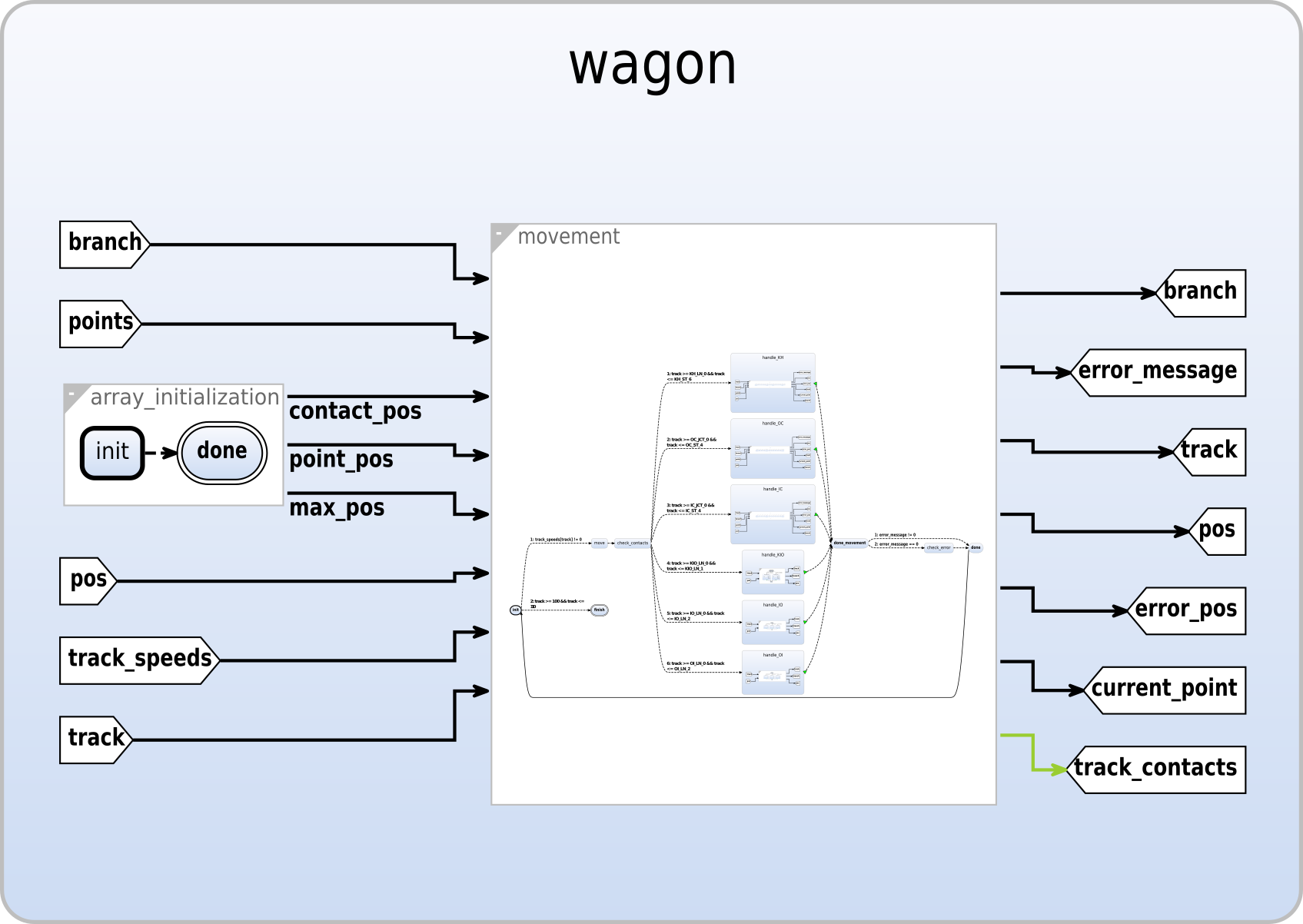}
		\caption{Top-down layout.}
		\label{fig:intro-wagon-top-down}
	\end{subfigure}
	\caption{Comparison of bottom-up and top-down layout for a compound graph. In this case, the graph is an SCChart that represents a railway wagon.}
	\label{fig:intro-wagon}
	\par\medskip
\end{figure*}

Automatic graph drawing is the process of creating a two-dimensional drawing of a graph.
There are many graph drawing algorithms for different purposes such as tree layout~\cite{DiBattistaETT94}, force- and stress-based layout~\cite{FruchtermanR91,GansnerKN05}, and layered layout~\cite{SugiyamaTT81}.
While most graph drawing algorithms operate on graphs where each node is an atomic element, the graph drawing problem can be expanded to include compound graphs where nodes can be parents of sub-graphs~\cite{EadesFL97,RufiangeMF12}.

A common approach to drawing compound graphs is to first draw the sub-graphs and then draw the parent nodes around them.
This recursive bottom-up approach provides a simple solution to determining sizes of nodes and it allows the entire graph to be laid out at the same scale.
The downside of the bottom-up approach is that drawings can become very large as the graphs grow in size.
A zoom-to-fit scaling then results in a scale where labels become illegible and also the overall structure is difficult to discern.
Another downside is that the overall variability in node sizes increases for graphs with deeper hierarchies due to different numbers of descendant nodes. This tends to result in large amounts of whitespace when these differently sized nodes are laid out~\cite{GutwengervHM+14}.

We propose a paradigm that aims to utilize computer screens effectively for drawing large compound graphs. The zoom-axis is treated as a third dimension for compound graph layouts using a top-down approach. The dimensions of parent nodes are fixed first and the child layouts are scaled down to fit within their parent. \autoref{fig:intro-wagon-top-down} shows an example of such a layout. In top-down layout the overall structure is discernible and the top-level labels are readable, facilitating systematic exploration of the whole model. As further motivation for the top-down layout approach, we consider the possibility of incremental layout. This has performance benefits both in terms of time and space. We can defer parts of the layout computation until we need them and we do not need to store the entire layout in memory immediately. This is especially useful for large models that are only partially visible at any given time.

The graph shown in \autoref{fig:intro-wagon} represents an SCChart~\cite{vonHanxledenDM+14}, which beyond nodes and edges also contains other elements such as different types of labels. While providing visualizations for languages such as SCCharts is the motivation for top-down layout, we consider the work presented here applicable to compound graphs in general.
There exist other graph layout approaches for large compound graphs that use top-down approaches~\cite{VehlowBW17,AbelloHK06,ArchambaultMA07a} but they are often more targeted at data visualization, which does not transfer well to the model-driven engineering domain that SCCharts belongs to.

\subsection*{Contributions and Outline}
\label{sec:contributions}
The contributions of this paper are the following:
\begin{itemize}
	\item We propose \emph{top-down layout}, a high-level, general-purpose layout framework for drawing compound graphs (\autoref{sec:topdown-layout}). Several methods for deciding node sizes are introduced in the form of \emph{node size approximators} and \emph{top-down aware layout algorithms}.
	\item Three possible configurations of top-down layout for SCCharts (\autoref{sec:sccharts}), practically validated with an implementation within the open-source \acs{elk}\footnote{\url{https://www.eclipse.org/elk/}} framework, demonstrating the advantages and disadvantages of top-down layout.
	\item Two quantitative evaluation metrics to reason about large two-dimensional graphs by modelling zoom as a third axis. These metrics are used to analyze and compare the different layout algorithms for a large set of SCCharts (\autoref{sec:evaluation}).
\end{itemize}

\section{\uppercase{Definitions}}
\label{sec:definitions}
\paragraph{Compound Graphs}
A \emph{compound graph} \(G\) consists of a set of nodes \(V\), a set of edges \(E\subseteq V\times V\) and a \emph{containment function} \(\tau':V\rightarrow 2^V\) that maps nodes to the set of their children.
This is the inverse of the \emph{hierarchy function} \(\tau:V\rightarrow V\) that is typically used in definitions of compound graphs~\cite{EadesFL97}, which maps nodes to their parents. For top-down layout we restrict edges to be between nodes that have the same parent, whereas additional considerations are necessary for \emph{hierarchy-crossing} edges.

A well-formed compound graph's containment function is a tree, which means that every tree can be mapped onto a compound graph, however not every compound graph can be mapped onto a tree. Trees are also sometimes called \emph{hierarchical graphs} and there are many different visualization techniques for them~\cite{EadesFL97,RufiangeMF12,DiBattistaETT94}.

A \emph{clustered graph} is structurally also a compound graph, but the term is typically used when the containment function is extracted from the underlying graph through a clustering method rather than being defined by the graph itself.

\paragraph{Compound Graph Layout}
A \emph{layout} or \emph{drawing} \(\Gamma\) of a compound graph is a mapping where each node is mapped to a bounding box and each edge is mapped to a curve in the plane. Positions are relative to the parent node if one exists.

A layout \(\Gamma\) of a compound node \(r\) is \emph{composed} of the layout \(\Gamma_r\) of its children and the respective layouts of each child's contents \(\Gamma_c\) together with a scale factor \(s\). This composition is expressed as \(\Gamma = \Gamma_r \circ (s,\{\Gamma_{c}|c\in \tau'(r)\})\).

Each layout \(\Gamma\) has a \emph{size} that is determined by the minimal bounding box around the nodes and edges included in \(\Gamma\). For a node \(v\) where \(\tau'(v) = \varnothing\), the size of its bounding box is a \emph{base size} defined by the node.

\section{\uppercase{Top-Down Layout}}
\label{sec:topdown-layout}
We now introduce \emph{top-down layout} as a general-purpose approach for laying out large compound graphs.
As explained, top-down layout aims to directly utilize zoom as a third dimension during layout to provide readable views of diagrams at any zoom level and to allow incremental graph layout.

\subsection{Algorithm}
Each node in a compound graph is assigned a \emph{node type} that controls how it is handled by the top-down layout procedure.

\begin{itemize}
	\item \textsc{Root} --- the entry point of top-down layout,
	\item \textsc{Flexible} --- nodes that are assigned a size and scale their children accordingly,
	\item \textsc{Fixed} --- nodes that do not scale their children.
\end{itemize}

\subsubsection{Layout Procedure}
In \Cref{alg:bottomup-layout} the traditional bottom-up layout process is shown. In order to invert the process to a top-down procedure, we switch the order of the \emph{recursive}, \emph{layout}, and \emph{size} steps and insert an additional \emph{scale} step. The resulting algorithm is shown in \Cref{alg:topdown-layout}. Here, we return the layout as a result in each step. This is for the sake of clarity, but it is not necessary. The layout can be stored directly on the graph in a single pass.

Additionally, we allow marking nodes for layout, which allows early termination and thus partial layout of a graph to save computation time. Inner nodes can be laid out later and composed into the already existing partial layout. The benefit is that layout calls can be deferred to when the result is actually required e.g., when we are zoomed in close enough to discern details. This \emph{incremental layout} is not possible using a bottom-up approach.

\begin{algorithm}
	\SetKwInOut{Input}{Input}\SetKwInOut{Output}{Output}
	\SetKw{And}{and}\SetKw{Or}{or}\SetKw{Is}{is}
	\DontPrintSemicolon
	
	\textsc{BottomUpLayout}($G,r$)\;
	\Input{Compound graph \(G = (V,E,\tau')\), current root node \(r\in V\)}
	\Output{Layout \(\Gamma\)}
	\BlankLine

	\tcp{Recursive call}
	\ForEach{$c\in \tau'(r)$}
	{
		$\Gamma_c \leftarrow \textsc{BottomUpLayout}(G,c)$\;
	}
	\tcp{Compute layout}
	$\Gamma_r \leftarrow layout(r)$\;

	\tcp{Set size according to layout}
	$r.size \leftarrow \Gamma_r.size$\;

	\tcp{Compose layouts for rendering}
	\Return{$ \Gamma_r \circ (1,\{\Gamma_{c}|c\in \tau'(r)\})$}
	
	\par\medskip
	\caption{General bottom-up layout procedure.}\label{alg:bottomup-layout}
\end{algorithm}

\begin{algorithm}
	\SetKwInOut{Input}{Input}\SetKwInOut{Output}{Output}
	\SetKw{And}{and}\SetKw{Or}{or}\SetKw{Is}{is}
	\DontPrintSemicolon

	\textsc{TopDownLayout}($G,r$)\;
	\Input{Compound graph \(G = (V,E,\tau')\), current root node \(r\in V\)}
	\Output{Layout \(\Gamma\)}
	\BlankLine
	
	\tcp{Set node sizes}
	\If{$r$ \Is \textsc{Flexible} \Or $r$ \Is \textsc{Root}}
	{
		\ForEach{$c \in \tau'(r)$}
		{
			$c.size \leftarrow predictSize(c)$\;
		}
	}

	\tcp{Compute layout}
	$\Gamma_r \leftarrow layout(r)$\;
	
	\If{$r$ \Is \textsc{Root}}
	{
		$r.size \leftarrow \Gamma_r.size $\;	
	}

	\tcp{Compute scale factor}

	\If{$r$ \Is \textsc{Flexible}}
	{
		$s_r \leftarrow r.size / \Gamma_r.size$\; 

	}
	\tcp{Recursive call}

	\ForEach{$c \in \tau'(r)$}
	{	
		\If{$markedForLayout(c)$}
		{
			$\Gamma_c \leftarrow \textsc{TopDownLayout}(G,c)$\;
		}
	}

	\tcp{Compose layouts for rendering}
	\Return{$ \Gamma_r \circ (s_r,\{\Gamma_{c}|c\in \tau'(r)\}) $} 
	\par\medskip
	\caption{General top-down layout procedure.}\label{alg:topdown-layout}
\end{algorithm}

\subsubsection{Size Prediction}

The core issue of performing a top-down layout is the task of choosing appropriate sizes for nodes. In order to compute a layout, we require the sizes of all nodes to be known. In bottom-up layout this is solved by using the size of the already calculated child layout. In top-down layout we change the order of the operations. Therefore, we need to set a size first. We propose two strategies:

\paragraph{Node Size Approximation}
The first step moving away from bottom-up layout is to make all nodes \textsc{Flexible} i.e., nodes that may apply a scale to their children.
The simplest method is to assign the base size to each node. However, this may result in large scale differences between parent nodes and their children, which creates a lot of whitespace.

A size approximator \(\mathcal{P}\) estimates the width and height of a compound graph \(G\). The aim of a size approximator in general is to minimize the amount of additional whitespace that is introduced when scaling down a child layout, while providing appropriate amount of space corresponding to the complexity of the child layout.
We have already implicitly introduced the simplest size approximator, which is the one that always returns the base size. A more sophisticated, but still very basic approximator, counts the children of the root node in \(G\) and takes the square root of that count as a multiplication factor for the base size. This is the default size approximator used in the later examples and is called the \emph{node count approximator}.

If a layout is produced that has an aspect ratio that is close to the aspect ratio of the base size, this heuristic tends to introduce little additional whitespace when compositing the child drawing into its parent.

\paragraph{Top-down Aware Layout Algorithms}
This approach uses \textsc{Fixed} nodes, which are container nodes for several nodes of the base size. The base size is a constant size defined by a diagram designer.
We use layout algorithms that can predict the size of their resulting layout before computing the layout.

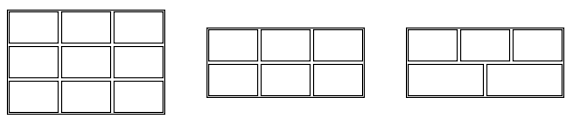
\begin{figure}[h]
	\centering
	\begin{subfigure}{.3\linewidth}
		\centering
		\begin{tikzpicture}[scale=.23]
			\draw (0.1,0.1) rectangle (2.9,1.9);
			\draw (3.1,0.1) rectangle (5.9,1.9);
			\draw (6.1,0.1) rectangle (8.9,1.9);

			\draw (0.1,2.1) rectangle (2.9,3.9);
			\draw (3.1,2.1) rectangle (5.9,3.9);
			\draw (6.1,2.1) rectangle (8.9,3.9);

			\draw (0.1,4.1) rectangle (2.9,5.9);
			\draw (3.1,4.1) rectangle (5.9,5.9);
			\draw (6.1,4.1) rectangle (8.9,5.9);

			\draw (0, 0) rectangle (9, 6);
		\end{tikzpicture}
		\caption{9 nodes can completely fill a \(3\times 3\) grid.}
		\label{fig:topdownpacking-1}
	\end{subfigure}
	\hfill
	\begin{subfigure}{.3\linewidth}
		\centering
		\begin{tikzpicture}[scale=.23]
			\draw (0.1,1.1) rectangle (2.9,2.9);
			\draw (3.1,1.1) rectangle (5.9,2.9);
			\draw (6.1,1.1) rectangle (8.9,2.9);

			\draw (0.1,3.1) rectangle (2.9,4.9);
			\draw (3.1,3.1) rectangle (5.9,4.9);
			\draw (6.1,3.1) rectangle (8.9,4.9);

			\draw (0, 1) rectangle (9, 5);
			\path (0, 0) rectangle (9, 6);
		\end{tikzpicture}
		\caption{For 6 nodes the final row is removed.}
		\label{fig:topdownpacking-2}
	\end{subfigure}
	\hfill
	\begin{subfigure}{.3\linewidth}
		\centering
		\begin{tikzpicture}[scale=.23]
			\draw (0.1,1.1) rectangle (4.4,2.9);
			\draw (4.6,1.1) rectangle (8.9,2.9);

			\draw (0.1,3.1) rectangle (2.9,4.9);
			\draw (3.1,3.1) rectangle (5.9,4.9);
			\draw (6.1,3.1) rectangle (8.9,4.9);

			\draw (0, 1) rectangle (9, 5);
			\path (0, 0) rectangle (9, 6);
		\end{tikzpicture}
		\caption{For 5 nodes the incomplete row is expanded.}
		\label{fig:topdownpacking-3}
	\end{subfigure}
	\caption{Different cases of Topdownpacking.}
	\label{fig:topdownpacking}
\end{figure}

\emph{Topdownpacking} is a simple layout algorithm that fulfils this property. It takes several nodes, gives each the base size, and arranges the nodes in a square grid. If \(n\) is the number of nodes and there is no integer solution to \(\sqrt{n}\), the grid will be incomplete.
The different cases are illustrated in \autoref{fig:topdownpacking}.
If an entire row is empty, it is removed.
Alternatively, it is possible to expand nodes vertically to fill the space and maintain the aspect ratio of the parent.
Afterwards, the nodes of the last row are expanded horizontally to fully use the available space provided by the parent node. 

\subsubsection{Hierarchy-crossing Edges}
\label{sec:hierarchical-edges}
Although top-down layout cannot directly deal with hierarchy-crossing edges. There are still techniques to realize top-down diagrams with hierarchy-crossing edges.
There are two basic approaches. The first technique is to add in the hierarchy-crossing edges after the layout is complete. For some applications simple straight line edges can be sufficient but more complex problems can also be solved using standalone edge routing algorithms. The second technique can be applied during the main layout procedure. Instead of modeling the hierarchy-crossing edges as a single edge, we split them into multiple edges and connect them at the boundaries of the compound nodes using ports. The advantage is that we do not need an additional post-layout processor for hierarchical edges and can therefore keep the benefits of top-down layout such as iterative diagram generation. The downside is that the scaling of the different edge sections would end up being different, which depending on the application could be undesired.

\subsection{Application on SCCharts}
\label{sec:sccharts}
We apply top-down layout to the visualization of SCCharts~\cite{vonHanxledenDM+14}, which are a dialect of Statecharts~\cite{HarelG96}.
Syntactically, Statecharts are Higraphs~\cite{Harel98+}, which are a visualization formalism combining advantages of graph drawings and Venn diagrams~\cite{Venn80} to visualize set-theoretic properties.
Our compound graphs do not include partial containment of nodes, but we can handle the \emph{partitioning} of nodes into what we call \emph{regions}.
Regions are modeled as regular nodes contained within a parent with the restriction that they cannot have any incident edges. 

SCCharts are a modeling language whose models are edited and viewed both textually and graphically.
The corresponding diagram is then synthesized and automatically laid out using \acs{elk}~\cite{SchneiderSvH13}.
SCCharts are synthesized into compound graphs with two semantically different types of nodes called \emph{states} and \emph{regions} as introduced above.
States are connected by edges.
A state can also have internal structure, in which case it is called a \emph{superstate}.
Superstates contain regions and regions always contain states.
The layout of states in a region is performed using a layered approach~\cite{SugiyamaTT81}.
The layout of regions in a superstate is performed using \emph{Rectpacking}~\cite{DomroesLvHJ21}.

\begin{figure*}[t]
	\begin{subfigure}{0.49\linewidth}
		\centering
		\includegraphics[width=\linewidth]{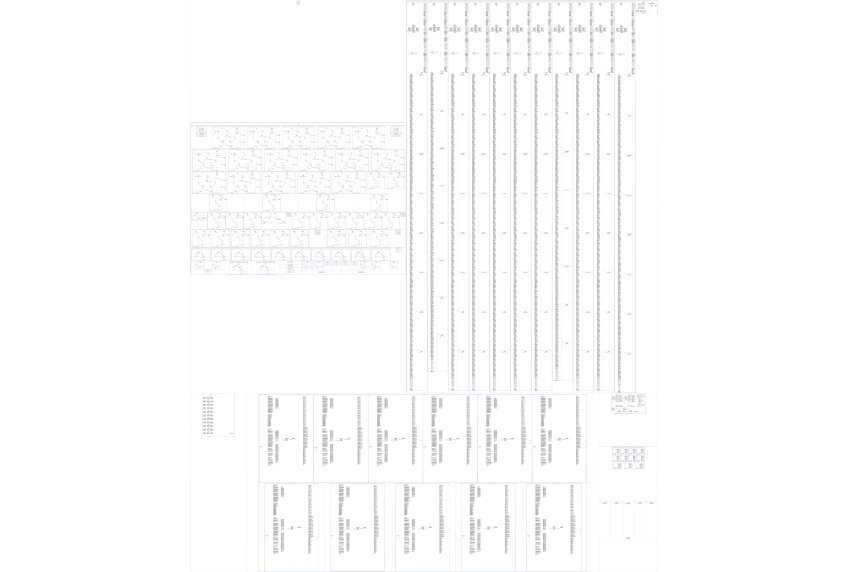}
		\caption{Bottom-up layout. All nodes are drawn at the same scale resulting in illegible labels at all levels.}
		\label{fig:bottom-up}
	\end{subfigure}
	\hfill
	\begin{subfigure}{0.49\linewidth}
		\centering
		\includegraphics[width=\linewidth]{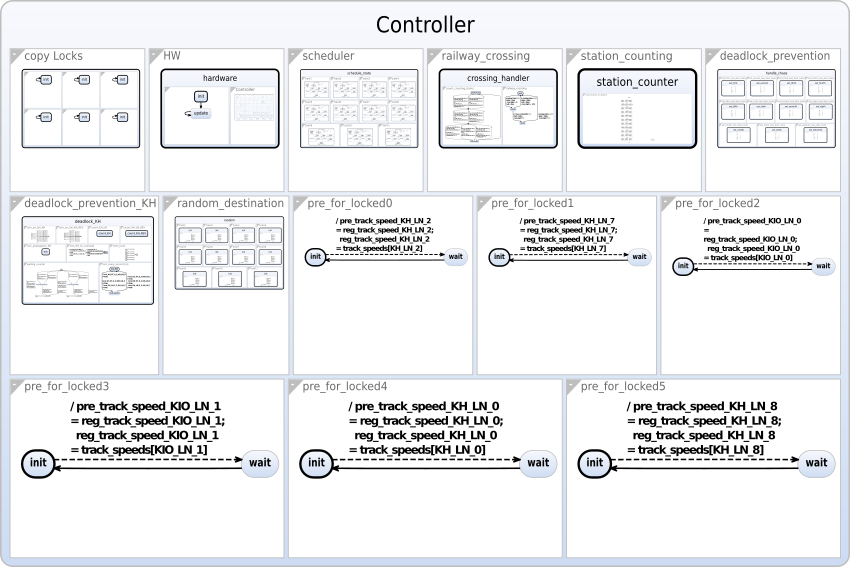}
		\caption{Top-down layout using \textsc{Flexible} states and regions. The node count approximator is used for all nodes.}
		\label{fig:top-down-nodecount}
	\end{subfigure}
	\par\medskip
	\begin{subfigure}{0.49\linewidth}
		\centering
		\includegraphics[width=\linewidth]{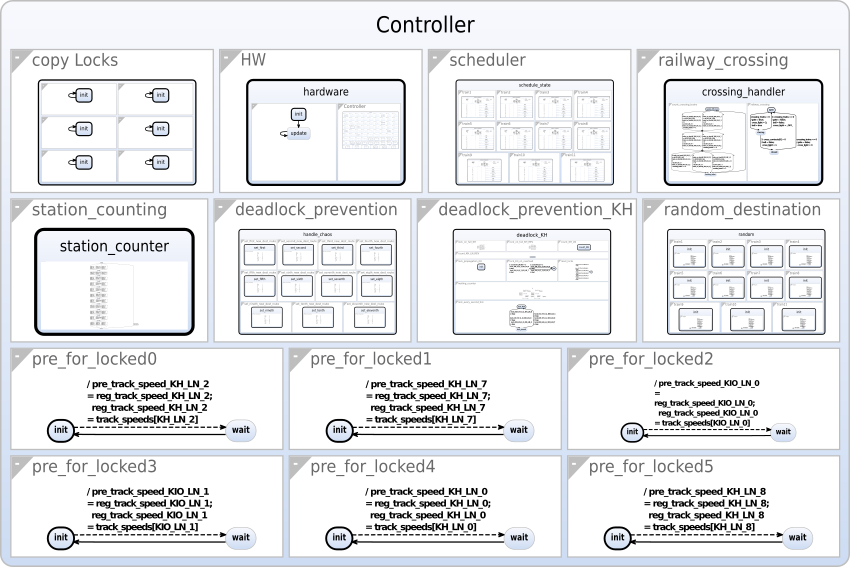}
		\caption{Top-down layout using \textsc{Flexible} states and regions. Look-ahead layout approximates the aspect ratios of regions.}
		\label{fig:top-down-lookahead}
	\end{subfigure}
	\hfill
	\begin{subfigure}{0.49\linewidth}
		\centering
		\includegraphics[width=\linewidth]{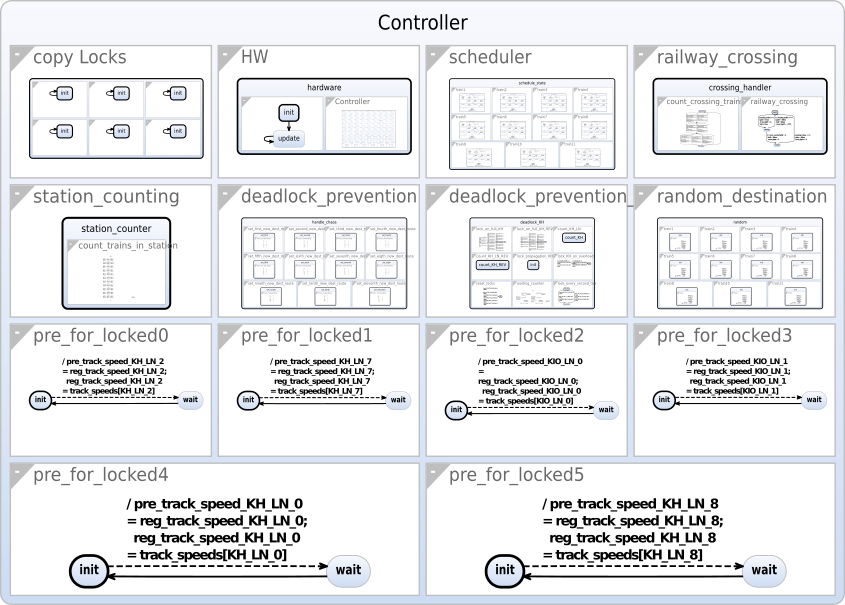}
		\caption{Top-down layout with \textsc{Fixed} states laid out by Topdownpacking and \textsc{Flexible} regions in a layered layout.}
		\label{fig:top-down-HP}
	\end{subfigure}
	\par\medskip
	\caption{Comparison of the bottom-up layout and three different top-down layout configurations for the \textsf{Controller} example SCChart.}
	\label{fig:top-down-comparison}
\end{figure*}

\subsubsection{Layout Configuration}
\label{sec:topdown-layout-variants}
We introduce three variants of top-down layout for SCCharts.
To illustrate the differences of these variants we use the \textsf{Controller} SCChart. This is an example from the railway domain, consisting of over 1700 states with a maximum hierarchical depth of 9. The textual source contains over 35\,000 lines. This is an example of a large model that may be browsed to understand the system architecture and the interactions of different components.

\autoref{fig:bottom-up} shows the traditional bottom-up drawing of \textsf{Controller}. This provides very little information to the observer. If one would scale \autoref{fig:bottom-up} to fit an A4 sheet, the largest font would have a size of 0.322 pt. In comparison, if one would scale \autoref{fig:top-down-HP} the same way, the title label font would be about 53 pt, and the font size of the top-level regions would be about 48 pt, which is quite readable.

\paragraph{Flexible Region sizes}
This first variant of top-down layout, shown in \autoref{fig:top-down-nodecount}, aims to be less invasive in the established visual style of SCCharts and to reduce scale discrepancies between regions. It is similar to the bottom-up procedure, but scaling may now be applied in each compound node. All nodes are \textsc{Flexible} and are laid out by their usual layout algorithm.

The essential step to produce aesthetic layouts is the node size approximation that is done for each child before computing the layout. Here we use the node count approximator. This results in states and regions having different sizes based on their contents, which reduces the difference in scale factors between sibling nodes while still maintaining a readability improvement.

\paragraph{Flexible Region sizes and Look-Ahead Layout}
This variant is shown in \autoref{fig:top-down-lookahead}.
To get an aspect ratio that fits the aspect ratio that we will get during the layout process, we can first compute the layout for one hierarchical layer and then take the resulting output dimensions and use them as our size approximation. If there are no grandchild nodes, this produces a perfect size prediction. In practice, we do not yet know how large the child nodes will be, because \textsc{Flexible} nodes are assigned sizes later.

During the approximation step, we set the sizes of the children using the node count approximator. For the case of SCCharts, this means that the look-ahead is suitable due to the alternation of regions and states. This is because the node count approximator works well in conjunction with Rectpacking, where we can set the target aspect ratio as an input.

\paragraph{Fixed Region Sizes}
\label{sec:H-P}
In this top-down variant for SCCharts, shown in \autoref{fig:top-down-HP}, we give all states the \textsc{Fixed} node type, while all regions are assigned the \textsc{Flexible} node type. The states in regions can be laid out using the layered approach. The resulting layout is then scaled down to fit the fixed region size. However, due to being \textsc{Fixed} nodes, the regions in states cannot be laid out using the normal Rectpacking algorithm that is used in bottom-up layout. Instead, we use Topdownpacking. The result is that all concurrent regions have the same sizes and can be easily viewed simultaneously. Their contents can have different scales, depending on the number of children a region has.

\section{\uppercase{Evaluation}}
\label{sec:evaluation}

This section introduces quantitative metrics for evaluating the quality of different diagrams regarding their utilization of the zoom axis of the drawing.

\subsection{Diagram Scale Normalization}
To accurately compare different diagrams, the scales used in them need to be normalized. For this, we introduce the \emph{z-level}, which represents the zoom level and the extension of a diagram into the third dimension.
More precisely, we need to define two fixed points that are used to normalize diagram scales allowing the comparison of distinct drawings at corresponding zoom levels.
We define our range of interest of \(z\) as the range \([0,1]\), where \(z=1\) means that the diagram is zoomed out to fit the entire drawing within a standard viewport (zoom-to-fit). In contrast, \(z=0\) means that we are zoomed in far enough that the smallest details are visible at their intended viewing size.
These boundaries define the points where there is no practical reason to zoom further in or out, respectively.

The choice of the viewport size is non-trivial because different display sizes enable different visualizations~\cite{JakobsenH13}. In our case, to work with some arbitrary but concrete numbers, we are examining a fixed viewport size with a resolution of 600 by 400 pixels. This emulates a diagram view as one window in a larger editor.

For the diagrams created and examined in the scope of this work, we have two distinct cases of scale usage in diagrams. Bottom-up diagrams start with a scale of 1 at \(z=0\) and build up the diagram around that. This means that the final zoomed-out diagram is drawn at a scale that is typically between 0 and 1, because the diagram needs to be scaled down to show everything at once unless it is very small.
For bottom-up diagrams, we define the diagram-specific constant \(a\) as the render scale of the diagram at \(z=1\).
In the case that diagrams are smaller than the viewport, \(a\) is greater than 1 and zoom-to-fit would enlarge the diagram until it filled the viewport. I.e., at \(z=1\) the diagram would be rendered at a scale larger than 1. In our analysis we set \(a\leq1\), i.e., stop enlarging the diagram once it is drawn at a scale of 1.

The other case is top-down layout. In this approach, the layout begins by giving the topmost node a scale of 1 and then, as it descends the hierarchy, nodes get smaller and smaller scales. This means that at \(z=1\) the render scale is 1, and that the diagram typically needs to be enlarged by some factor \(a > 1\) to show the smallest details. For top-down diagrams, \(a\) can be computed by finding the diagram element with the smallest scale and taking the inverse. This means that increasing the diagram drawing size by that factor \(a\) results in drawing the smallest element at its intended size. In other words, at \(z=0\), a top-down layout is drawn at scale \(a\).
In a nutshell, in bottom-up diagrams \(a\) is the minimum scale factor, and in top-down diagrams \(a\) is the maximum scale factor. To compare these two distinct linear scalings, we translate between diagram scales and \(z\)-level.

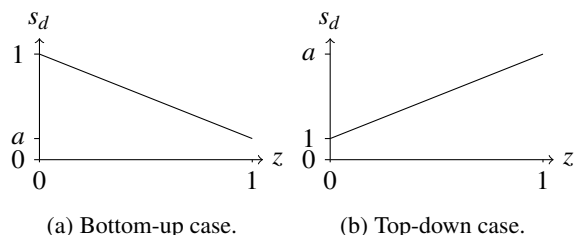
\begin{figure}
\centering
\begin{subfigure}{0.49\linewidth}
	\centering
	\begin{tikzpicture}[xscale=.7, yscale=.35]
		\draw[->] (0, 0) -- (4.2, 0) node[right] {\(z\)};
		\draw (0, 0) -- (0, -0.1) node[below] {\(0\)};
		\draw (4, 0) -- (4, -0.1) node[below] {\(1\)};

		\draw[->] (0, 0) -- (0, 4.6) node[above] {\(s_d\)};
		\draw (0, 0) -- (-0.1, 0) node[left] {\(0\)};
		\draw (0, 0.8) -- (-0.1, 0.8) node[left] {\(a\)};
		\draw (0, 4) -- (-0.1, 4) node[left] {\(1\)};

		\draw (0, 4) -- (4, 0.8);
	\end{tikzpicture}
	\caption{Bottom-up case.}
	\label{fig:scale-z-conversion-bottomup}
\end{subfigure}
\hfill
\begin{subfigure}{0.49\linewidth}
	\centering
	\begin{tikzpicture}[xscale=.7, yscale=.35]
		\draw[->] (0, 0) -- (4.2, 0) node[right] {\(z\)};
		\draw (0, 0) -- (0, -0.1) node[below] {\(0\)};
		\draw (4, 0) -- (4, -0.1) node[below] {\(1\)};

		\draw[->] (0, 0) -- (0, 4.6) node[above] {\(s_d\)};
		\draw (0, 0) -- (-0.1, 0) node[left] {\(0\)};
		\draw (0, 0.8) -- (-0.1, 0.8) node[left] {\(1\)};
		\draw (0, 4) -- (-0.1, 4) node[left] {\(a\)};

		\draw (0, 0.8) -- (4, 4);
	\end{tikzpicture}
	\caption{Top-down case.}
	\label{fig:scale-z-conversion-topdown}
\end{subfigure}
\caption{Conversion from \(z\)-level to diagram scale.}
\label{fig:scale-z-conversion}
\end{figure}

For both bottom-up and top-down layouts, we can determine \(a\) and use that value to compute the current diagram scale \(s_d\) for a given \(z\) using the following case distinction, illustrated in \autoref{fig:scale-z-conversion}.
This allows us to compare bottom-up and top-down layouts.

\begin{equation}
	s_d(a,z) = \begin{cases}
		z\cdot(a-1)+1, & \text{if } 0 < a \leq{1}\\
		z\cdot(1-a) + a, & \text{if } a > 1
	\end{cases}
\end{equation}

\subsection{Evaluation Metrics}
We now introduce two contrasting quantitative metrics. The \emph{readability}, which considers the legibility of texts measured across the zoom level, and the \emph{scale discrepancy}, which describes the scale differences between topologically close nodes.

\subsubsection{Readability}
\label{sec:readability}
Diagrams typically contain text in addition to graphical notation, i.e., nodes and edges.
It is difficult to quantitatively determine at what scale different drawings are readable in the sense that a viewer can understand their meanings.
For texts, this is a simpler task since we can consider the on-screen text size.
For example, at a font size of 12~pt text is easily legible for the average reader, but as it gets smaller, it becomes increasingly illegible.
Different font sizes may be used throughout the diagram, e.g., to distinguish node titles from other labels.
We abstract away from this by assuming the diagram designer has chosen suitable font sizes.

A given label \(t\) has a unit size. The text scale \(s_t\) is a multiplier that determines the actual on-screen size and is defined by the scale factor of the containing node. The diagram itself also has a scale \(s_d(z)\) that is 1 when the diagram is rendered at the size of its base elements. I.e., \(s_d(z)=1\) for \(z=1\) in top-down drawings and \(s_d(z)=1\) for \(z=0\) in bottom-up drawings.
Base elements are those elements that have a text scale \(s_t=1\).
The \emph{render scale} \(rs\) of an element \(t\), i.e., the drawn on-screen scale, is calculated as \(rs_t(z) = s_t\cdot s_d(z)\), which is simply the multiplication of the scale factor and the current zoom level of the diagram.

\begin{figure*}
	\centering
	\includegraphics[width=\linewidth]{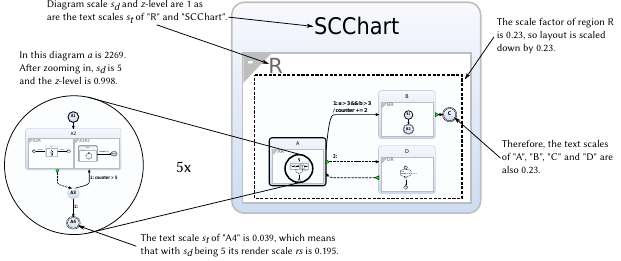}
	\caption{Illustration of the scale calculations that are done in top-down layouts.}
	\label{fig:topdown-calculations-demo}
\end{figure*}

In top-down layouts children are always scaled relative to their parent. This means that although the \(z\)-level is a linear scale, the text scales decrease exponentially as we descend the hierarchy.
Examples for concrete scale values are shown in \autoref{fig:topdown-calculations-demo}.

We consider elements \emph{readable} for a given \(z\)-level if \(rs_t(z) \geq 1\) holds. Let \(\mathcal{R}(z)\) be the set of readable texts at zoom level \(z\), and let \(\mathcal{T}\) be the set of all texts in a diagram. The fraction of readable texts is \(r(z) = |\mathcal{R}(z)| / |\mathcal{T}|\).
As we zoom into a diagram, the fraction of the diagram included in the viewport shrinks quadratically. We assume for simplicity that the aspect ratios of the viewport and the diagram are equal. This corresponds to the act of zooming out until the entire diagram fits into the viewport. We denote the area of the viewport as \(A_v\) and the area of the diagram as \(A_d\) to obtain the proportion \(v(z)\) of the diagram visible in the current viewport.

\begin{equation}
	v(z) = \textrm{min}\left(\frac{A_v}{A_d \cdot s_d(z)^2}, 1\right)
\end{equation}
\label{eq:viewport-proportion}

The quadratic term stems from the fact that both width and height are scaled. \(v(z)\) must be capped to a maximum of 1, as larger values only mean that the diagram is smaller than the viewport and could be further enlarged to fit the whole viewport. This does not increase the number of onscreen elements any further. We model the \emph{readability} of the diagram as a combination of the proportions of readable and visible texts.

\begin{equation}
	R(z) = r(z) \cdot v(z)
\end{equation}

The aim is to spread the readability out uniformly across \(z\) so that at any given zoom level there is always something readable.
We can see this in \autoref{fig:top-down-comparison} and \autoref{fig:topdown-calculations-demo}. In \autoref{fig:bottom-up} it is impossible to read anything, while in the various top-down layouts there is always some text at a readable scale.

\subsubsection{Scale Discrepancy}
\label{sec:scale-discrepancy}
In a traditional bottom-up layout, all parts of the diagram are drawn at the same scale. This means that when panning across the diagram, the zoom level does not need to be adjusted to view different parts of the drawing. Several areas can be viewed simultaneously, with the caveat that they do still need to be close to each other and not too large.

In a diagram laid out using a top-down approach, nodes may have different scales, and these scale discrepancies can be large enough that it is difficult to view both nodes at the same time without adjusting the zoom level.
Each node \(n\) in a compound graph has an associated scale \(s_n\) that defines the scale that is applied to the node's child layout. The \emph{scale discrepancy} \(D(n)\) is defined as

\begin{equation}
	D(n) = \frac{\textrm{max}(\mathcal{S}_n)}{\textrm{min}(\mathcal{S}_n)} - 1,
\end{equation}

where \(\mathcal{S}_n\) is the set of scale factors of the children of \(n\).
The discrepancy then describes how much smaller the smallest scale is relative to the largest scale. The scale discrepancy is 0 if all scales are equal.
Scale discrepancy is only examined on a topologically local level. This means that the nodes that are being compared have at most one intermediate node on a path between them. These paths can contain edges and child-parent relations.
Nodes that are close together are often viewed at the same time. They should be legible at the same \(z\)-level.
Nodes that are not close to each other are usually not viewed simultaneously, and therefore their discrepancy is of lower priority.

\subsection{Applying the Metrics}
We apply the introduced \emph{readability} and \emph{scale discrepancy} metrics on a large set of SCCharts. These stem from industrial projects and from class assignments. Our goal is to showcase the usefulness of the metrics in reasoning about the efficacy of different top-down layout variants. Additionally, we demonstrate the difference to standard bottom-up diagrams that do not utilize scaling.
\begin{figure}[h]
	\centering
	\includegraphics[width=\linewidth]{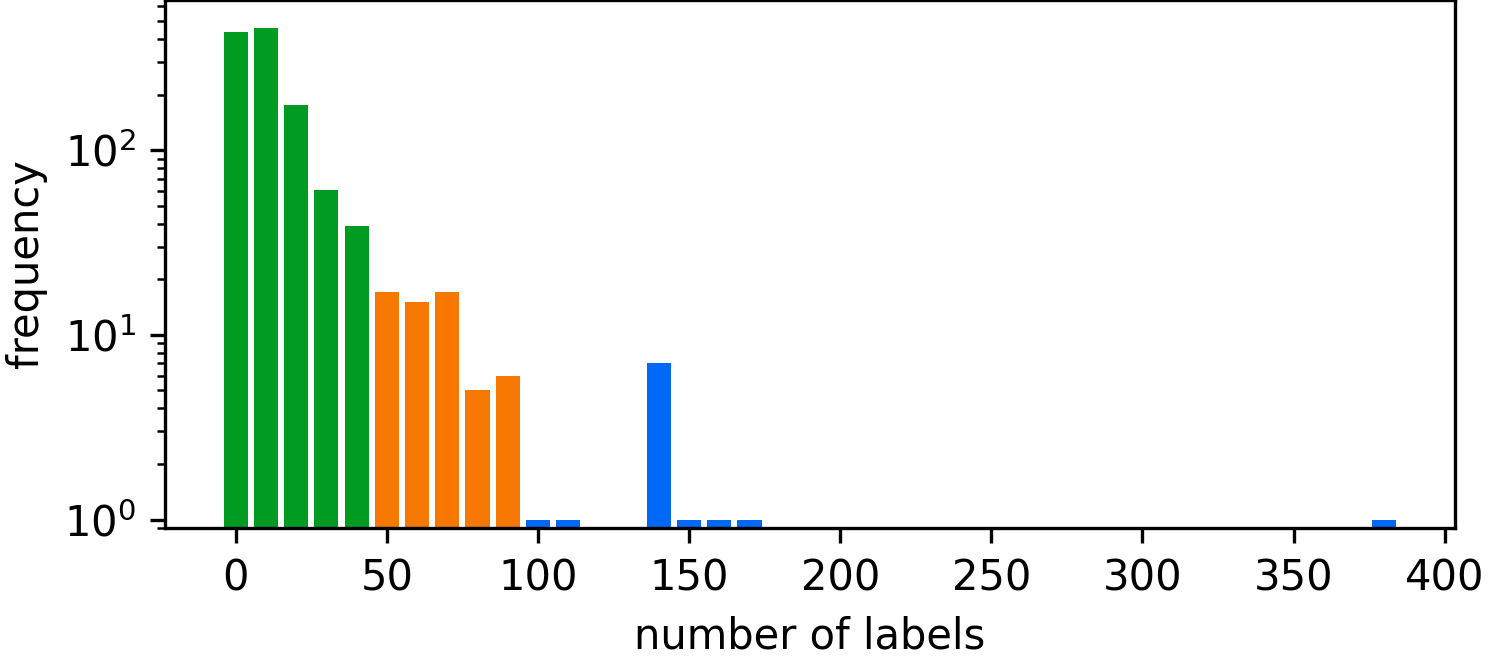}
	\caption{Distribution of the graph sizes according to the number of labels.}
	\label{fig:eval-size-distribution}
\end{figure}

The examined diagrams are generated from a set of 1250 SCCharts. The distribution of the models' size variation is captured in \autoref{fig:eval-size-distribution}. We group the dataset into \textcolor{small}{small}, \textcolor{medium}{medium} and \textcolor{large}{large} models to compare the effect model size has on the different algorithms. We use the number of labels as the size measure, because the metrics focus on the legibility of diagrams. There are also diagrams with many labels that end up being small, and large diagrams with few labels.
\begin{figure*}[t]
	\centering
	\begin{subfigure}{0.48\linewidth}
		\centering
		\includegraphics[width=\linewidth]{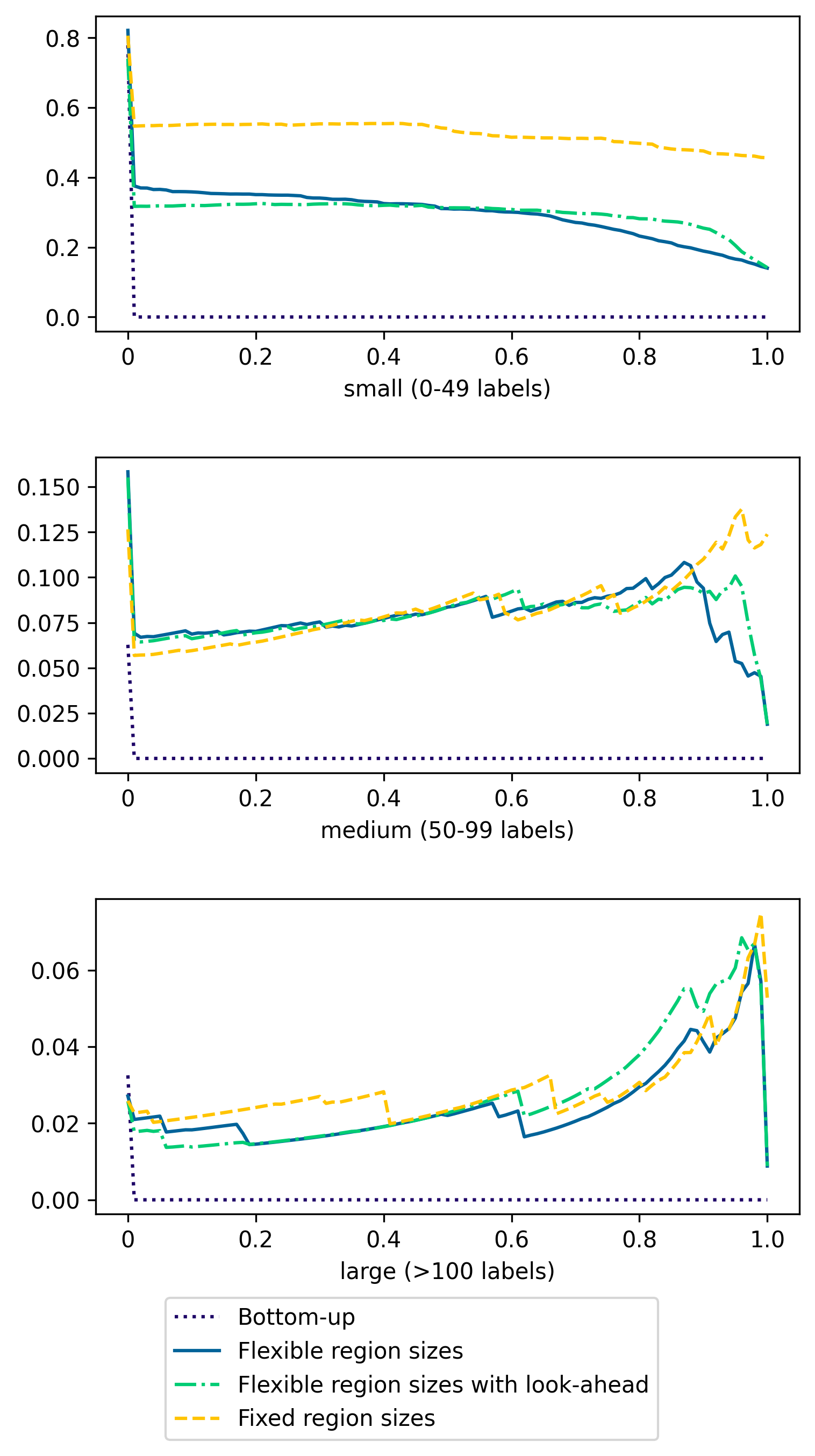}
		\caption{Average readabilities across \(z\)-level (higher is better).}
		\label{fig:eval-avg-readability}
	\end{subfigure}
	\hfill
	\begin{subfigure}{0.48\linewidth}
		\centering
		\includegraphics[width=\linewidth]{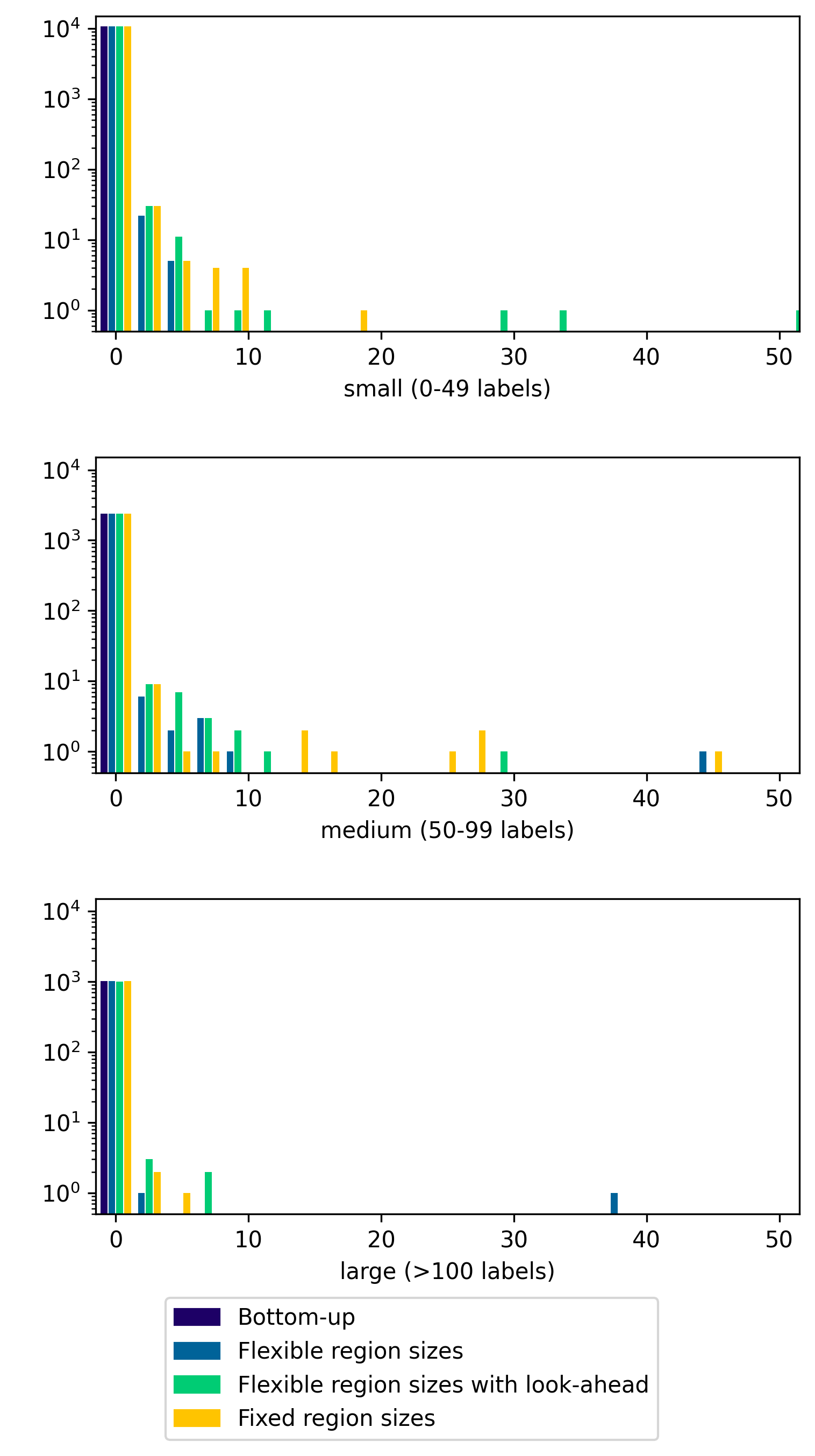}
		\caption{Histograms of scale discrepancies (further left is better).}
		\label{fig:eval-avg-local-scale-discrepancy}
	\end{subfigure}
	\par\medskip
	\caption{Evaluation of readability and scale discrepancy across 1250 SCCharts. Note that the vertical scales in \autoref{fig:eval-avg-readability} differ.}
	\label{fig:metric-evaluation}
\end{figure*}

For readability and scale discrepancy, measurements were taken for all of the sample diagrams. The established bottom-up layout and all three top-down layout variants outlined in \autoref{sec:topdown-layout-variants} were examined. 

As previously discussed, good average readability is characterized by being approximately uniformly distributed across the entire \(z\) range. This would mean that no matter at what zoom level a diagram is being viewed, the proportion of readable elements in the viewport would remain approximately constant. As diagrams become larger, that ideal constant decreases, because the zoom range in which the labels are distributed becomes larger.
Scale discrepancy should be minimized. Bottom-up layout naturally has no scale discrepancy, but for top-down layout we can compare how well the different variants minimize scale discrepancy in differently sized graphs.

\autoref{fig:eval-avg-readability} shows the readability metric for all algorithms grouped by the three size groups.
For bottom-up layout readability is always 0 until the legibility threshold is reached. The smallest diagrams reach an average readability of about 0.75 at the lowest \(z\)-level. This indicates that on average, three-quarters of the smallest diagrams fit into the viewport. For the larger groups, this value is a lot smaller. 
For the top-down variants, we can observe that in general, they get close to uniform readability.
At the highest \(z\)-level the readability sometimes drops sharply. This could be because many diagrams have one readable title and many small children. As we zoom further in, these large discrepancies are quickly reduced.
Further observations are that the fixed region sizes algorithm performs better for smaller diagrams, but both flexible scale algorithms appear to produce better results for the larger diagrams. Look-ahead layout produces the most consistent readabilities for small and medium graphs.
It is important to keep in mind that the readability for the different groups of graphs is on different vertical scales. For small diagrams the readability reaches up to 0.8, whereas for large diagrams the highest reached readability is only about 0.07. This is to be expected since as the number of labels goes up, the overall readability decreases.

In \autoref{fig:eval-avg-local-scale-discrepancy} the histogram shows how the scale discrepancies are distributed in the examined dataset for each algorithm. The histograms are restricted to scale discrepancies between 0 and 50, but notably, for the flexible scale look-ahead layout algorithm, there are some extreme outliers in the area of around 200.
There is no scaling in bottom-up layout, so there are no scale discrepancies. Both flexible scale algorithms produce consistently smaller scale discrepancies than the fixed scale method.

\section{\uppercase{Related Work}}
\label{sec:related}
We focus on abstract visualization techniques for compound graphs and on the effects of scale, both in the underlying data and the resulting visualizations.

\paragraph{Visualizing Trees}
In approaches where all elements of different hierarchies are treated equally, the resulting diagrams grow rapidly in their space requirements. A concept to handle this is a \emph{fractal layout}, where the diagram is self-similar. This means that no matter which hierarchy level is being viewed, the space requirements of what is shown in the viewing area are always similar~\cite{KoikeY93}. Top-down layout also aims to achieve fractal self-similarity with a general-purpose approach that can be applied to any model containing hierarchical structure.

Balloon trees and their variations~\cite{Holten06,MelanconH98,CarriereK95} are examples of self-similar diagram types. Tree graphs are restructured into compound graphs where the child relation is restricted to one layer for an algorithm such as radial tree layout, and the grandchildren are contained within their parents.
The top-down layout presented here, can be combined with Eades's radial layout~\cite{Eades92} to produce balloon tree-like drawings.

Treemap layouts are an approach for drawing hierarchical graphs that aim to fill the available space effectively. Each node in a treemap receives an area proportional to its weight.
Typically, square aspect ratios of the nodes are a desired goal~\cite{ShneidermanW01}.

\paragraph{Multilevel Layout Approaches}
Top-down divide-and-conquer approaches to graph layout have previously been used to tackle large graphs in combination with force-directed layout algorithms~\cite{Walshaw00,GajerGK00,HachulJ05,BourquiAM07}. By first clustering groups of vertices, a computationally expensive algorithm can be applied to a smaller graph. This process can then be repeated on the individual clusters until eventually all vertices have been laid out.
However, this is not applicable if the hierarchical structure is already given, as is the case for SCCharts.

Group-in-Box layouts also deal with clustered multilevel approaches but additionally use a treemap layout to arrange the clusters in boxes~\cite{RodriguesMSS+11,ChaturvediDAZ+14}. We also combine different layout algorithms in our multilevel approach but in the concrete application the choice of algorithms is left up to the diagram designer.

\paragraph{Geographic Map Exploration}
Common map and navigation apps, such as Google Maps, also face the problem of visualizing (geographic) data sets that are much too large to be displayed all at once.
The main technique to solve that problem is \emph{information filtering}.
That approach has inspired several works to visualize large graphs, for example by aggregating nodes and edges, as illustrated for example by GraphMaps~\cite{MondalN18}.

We share the goal of making large graphs explorable by just zooming and panning.
This includes the aim to show (at least some) labels at legible size irrespective of the current zoom level.
Our approach differs in that we build on graphs that are already hierarchically structured, and that we tackle the problem by varying scales.
However, a natural extension of our approach is to also filter out graph elements that are below a certain size.
Such filtering, which actually is already part of our tooling but is not discussed further here, also improves rendering performance.

\paragraph{Zoomable User Interfaces}
A \ac{zui} is a paradigm that differs from the typical window and desktop metaphor.
Instead of a bounded window, the interface is conceptually an infinite plane where user interface elements can be freely positioned and scaled~\cite{BedersonH94}.
This adds additional freedom in interface design but also comes with challenges, especially regarding user interaction.
Bedersen et al.\ developed a general framework for creating \ac{zui}s~\cite{BedersonMG00}, and Good and Bederson demonstrated its application for slide show presentations~\cite{GoodB02}.
Top-down layout uses the same principles to display information in a compact and explorable manner.

\paragraph{Zooming and Layout}
Many techniques for improving the navigation and readability of large diagrams leverage capabilities available by viewing diagrams on computer screens. The most basic technique is \emph{pan+zoom}, which allows us to navigate large graphs on comparatively small screens. Other approaches are more invasive, actively distorting or dynamically adjusting the drawing. These \emph{focus+context} techniques include fisheye lenses~\cite{SarkarB92} and dynamic expansion of hierarchies. This expansion is also a standard diagram interaction method of the \ac{klighd} framework~\cite{SchneiderSvH13} used by SCCharts~\cite{vonHanxledenDM+14}.

While these techniques improve the navigability of large diagrams, they do not utilize the zoom capability at the layout step of the graph visualization. We emulate zooming during graph layout by scaling parts of the drawing down.
Top-down layout enhances the usability of pan+zoom by utilizing scale.

In this paper we focus on a static approach to top-down layout. We do not consider dynamic approaches such as semantic zooming~\cite{BedersonH94}. However, we believe both approaches can and should be combined to provide the best possible user experience.

\section{\uppercase{Conclusion}}
\label{sec:conclusion}
We introduced a general-purpose layout approach to construct top-down diagrams with arbitrary layout algorithms and a large degree of configuration flexibility. We also established a general framework for analyzing and discussing drawings of compound diagrams that actively use scale as part of their visualizations.
We put the designed metrics into practice by analyzing several variants of top-down layout, and the expectations lined up with the results.

One goal is to find a configuration and methods of node size approximation that allow a good balance of readability and scale discrepancy, and a qualitative evaluation with user feedback in addition to quantitative metrics. An ideal result would be a layout that looks like a bottom-up layout at the microscale, but like a top-down layout at the macroscale. Good size approximation is the key to achieve this. 

Top-down layout is a promising approach for handling large compound graphs and opens up many applications. Overall, the top-down approach results in diagrams that are more compact drawings of large graphs with a better general overview at each hierarchical level than the bottom-up counterpart.
With SCCharts as a concrete application for top-down layout, we demonstrated that the principles can be applied to a wide range of similar domains.

Future work will focus on improved node size approximations that are both effective and efficient.
Additionally, further applications of top-down layout are incremental and partial layout of large, potentially unbounded, graphs as well as the combination of our static layout approach with dynamic filtering and techniques as used in common map viewing tools.

\bibliographystyle{apalike}
{\small
\bibliography{../../bib/cau-rt,../../bib/pub-rts.bib,../../bib/rts-arbeiten.bib}}

\begin{thebibliography}{}

\bibitem[Abello et~al., 2006]{AbelloHK06}
Abello, J., van Ham, F., and Krishnan, N. (2006).
\newblock Ask-graphview: {A} large scale graph visualization system.
\newblock {\em {IEEE} Trans. Vis. Comput. Graph.}, 12(5):669--676.

\bibitem[Archambault et~al., 2007]{ArchambaultMA07a}
Archambault, D., Munzner, T., and Auber, D. (2007).
\newblock Topolayout: Multilevel graph layout by topological features.
\newblock {\em IEEE Transactions on Visualization and Computer Graphics},
  13(2):305--317.

\bibitem[Bederson and Hollan, 1994]{BedersonH94}
Bederson, B.~B. and Hollan, J.~D. (1994).
\newblock Pad++: {A} zooming graphical interface for exploring alternate
  interface physics.
\newblock In {\em {UIST} '94: Proceedings of the 7th annual {ACM} symposium on
  User interface software and technology}, pages 17--26, New York, NY, USA. ACM
  Press.

\bibitem[Bederson et~al., 2000]{BedersonMG00}
Bederson, B.~B., Meyer, J., and Good, L. (2000).
\newblock Jazz: an extensible zoomable user interface graphics toolkit in java.
\newblock In {\em Proceedings of the 13th annual ACM symposium on User
  interface software and technology}, pages 171--180.

\bibitem[Bourqui et~al., 2007]{BourquiAM07}
Bourqui, R., Auber, D., and Mary, P. (2007).
\newblock How to draw clusteredweighted graphs using a multilevel
  force-directed graph drawing algorithm.
\newblock In {\em Information Visualization, 2007. IV '07. 11th International
  Conference}, pages 757--764.

\bibitem[Carriere and Kazman, 1995]{CarriereK95}
Carriere, J. and Kazman, R. (1995).
\newblock Interacting with huge hierarchies: Beyond cone trees.
\newblock In {\em Proceedings of Visualization 1995 Conference}, pages 74--81.
  IEEE.

\bibitem[Chaturvedi et~al., 2014]{ChaturvediDAZ+14}
Chaturvedi, S., Dunne, C., Ashktorab, Z., Zachariah, R., and Shneiderman, B.
  (2014).
\newblock Group-in-a-box meta-layouts for topological clusters and
  attribute-based groups: Space-efficient visualizations of network communities
  and their ties.
\newblock In {\em Computer Graphics Forum}, volume~33, pages 52--68. Wiley
  Online Library.

\bibitem[{Di Battista} et~al., 1994]{DiBattistaETT94}
{Di Battista}, G., Eades, P., Tamassia, R., and Tollis, I.~G. (1994).
\newblock Algorithms for drawing graphs: An annotated bibliography.
\newblock {\em Computational Geometry: Theory and Applications}, 4:235--282.

\bibitem[Domr{\"o}s et~al., 2021]{DomroesLvHJ21}
Domr{\"o}s, S., Lucas, D., von Hanxleden, R., and Jansen, K. (2021).
\newblock On order-preserving, gap-avoiding rectangle packing.
\newblock In {\em Proceedings of the 16th International Joint Conference on
  Computer Vision, Imaging and Computer Graphics Theory and Applications
  (VISIGRAPP 2021) - Volume 3: IVAPP}, pages 38--49. INSTICC, SciTePress.

\bibitem[Eades, 1992]{Eades92}
Eades, P. (1992).
\newblock Drawing free trees.
\newblock In {\em Bulletin of the Institute for Combinatorics and its
  Applications}, volume~5, pages 10--36.

\bibitem[Eades et~al., 1997]{EadesFL97}
Eades, P., Feng, Q.-W., and Lin, X. (1997).
\newblock Straight-line drawing algorithms for hierarchical graphs and
  clustered graphs.
\newblock In North, S., editor, {\em Graph Drawing}, volume 1190 of {\em LNCS},
  pages 113--128. Springer Berlin / Heidelberg.

\bibitem[Fruchterman and Reingold, 1991]{FruchtermanR91}
Fruchterman, T. M.~J. and Reingold, E.~M. (1991).
\newblock Graph drawing by force-directed placement.
\newblock {\em Software---Practice \& Experience}, 21(11):1129--1164.

\bibitem[Gajer et~al., 2000]{GajerGK00}
Gajer, P., Goodrich, M.~T., and Kobourov, S.~G. (2000).
\newblock A fast multi-dimensional algorithm for drawing large graphs.
\newblock In {\em Graph Drawing’00 Conference Proceedings}, pages 211--221.

\bibitem[Gansner et~al., 2005]{GansnerKN05}
Gansner, E.~R., Koren, Y., and North, S.~C. (2005).
\newblock Graph drawing by stress majorization.
\newblock In Pach, J., editor, {\em Graph Drawing}, volume 3383 of {\em LNCS}.
  Springer Berlin Heidelberg.

\bibitem[Good and Bederson, 2002]{GoodB02}
Good, L. and Bederson, B.~B. (2002).
\newblock Zoomable user interfaces as a medium for slide show presentations.
\newblock {\em Information Visualization}, 1(1):35--49.

\bibitem[Gutwenger et~al., 2014]{GutwengervHM+14}
Gutwenger, C., von Hanxleden, R., Mutzel, P., R{\"u}egg, U., and Sp{\"o}nemann,
  M. (2014).
\newblock Examining the compactness of automatic layout algorithms for
  practical diagrams.
\newblock In {\em Proceedings of the Workshop on Graph Visualization in
  Practice (GraphViP '14)}, pages 42--52.

\bibitem[Hachul and J{\"u}nger, 2005]{HachulJ05}
Hachul, S. and J{\"u}nger, M. (2005).
\newblock Drawing large graphs with a potential-field-based multilevel
  algorithm.
\newblock In {\em Revised Selected Papers of the 12th International Symposium
  on Graph Drawing (GD '04)}, volume 3383 of {\em LNCS}, pages 285--295.
  Springer.

\bibitem[Harel, 1988]{Harel98+}
Harel, D. (1988).
\newblock On visual formalisms.
\newblock {\em Commun. ACM}, 31(5):514–530.

\bibitem[Harel and Gery, 1996]{HarelG96}
Harel, D. and Gery, E. (1996).
\newblock Executable object modeling with statecharts.
\newblock In {\em Proceedings of the 18th International Conference on Software
  Engineering}, ICSE '96, pages 246--257. IEEE Computer Society.

\bibitem[Holten, 2006]{Holten06}
Holten, D. (2006).
\newblock Hierarchical edge bundles: Visualization of adjacency relations in
  hierarchical data.
\newblock {\em IEEE Transactions on Visualization and Computer Graphics},
  12(5):741--748.

\bibitem[Jakobsen and Hornbæk, 2013]{JakobsenH13}
Jakobsen, M. and Hornbæk, K. (2013).
\newblock Interactive visualizations on large and small displays: The
  interrelation of display size, information space, and scale.
\newblock {\em IEEE transactions on Visualization and Computer Graphics},
  19:2336--45.

\bibitem[Koike and Yoshihara, 1993]{KoikeY93}
Koike, H. and Yoshihara, H. (1993).
\newblock Fractal approaches for visualizing huge hierarchies.
\newblock In Glinert, E.~P. and Olsen, K.~A., editors, {\em Proc. {IEEE} Symp.
  Visual Languages, {VL}}, pages 55--60. IEEE Computer Society.

\bibitem[Melancon and Herman, 1998]{MelanconH98}
Melancon, G. and Herman, I. (1998).
\newblock Circular drawings of rooted trees.
\newblock {\em Information Systems [INS]}, (R 9817).

\bibitem[Mondal and Nachmanson, 2018]{MondalN18}
Mondal, D. and Nachmanson, L. (2018).
\newblock A new approach to graphmaps, a system browsing large graphs as
  interactive maps.
\newblock In Telea, A.~C., Kerren, A., and Braz, J., editors, {\em Proceedings
  of the 13th International Joint Conference on Computer Vision, Imaging and
  Computer Graphics Theory and Applications {(VISIGRAPP} 2018) -- Volume 3:
  IVAPP}, pages 108--119, Funchal, Madeira, Portugal. SciTePress.

\bibitem[Rodrigues et~al., 2011]{RodriguesMSS+11}
Rodrigues, E.~M., Milic-Frayling, N., Smith, M., Shneiderman, B., and Hansen,
  D. (2011).
\newblock Group-in-a-box layout for multi-faceted analysis of communities.
\newblock In {\em 2011 IEEE Third International Conference on Privacy,
  Security, Risk and Trust and 2011 IEEE Third International Conference on
  Social Computing}, pages 354--361. IEEE.

\bibitem[Rufiange et~al., 2012]{RufiangeMF12}
Rufiange, S., McGuffin, M.~J., and Fuhrman, C.~P. (2012).
\newblock Treematrix: A hybrid visualization of compound graphs.
\newblock {\em Computer Graphics Forum}, 31(1):89--101.

\bibitem[Sarkar and Brown, 1992]{SarkarB92}
Sarkar, M. and Brown, M.~H. (1992).
\newblock Graphical fisheye views of graphs.
\newblock In {\em Proceedings of the SIGCHI Conference on Human Factors in
  Computing Systems}, pages 83--91. ACM.

\bibitem[Schneider et~al., 2013]{SchneiderSvH13}
Schneider, C., Sp{\"o}nemann, M., and von Hanxleden, R. (2013).
\newblock Just model! -- {P}utting automatic synthesis of node-link-diagrams
  into practice.
\newblock In {\em Proceedings of the IEEE Symposium on Visual Languages and
  Human-Centric Computing (VL/HCC '13)}, pages 75--82, San Jose, CA, USA. IEEE.

\bibitem[Shneiderman and Wattenberg, 2001]{ShneidermanW01}
Shneiderman, B. and Wattenberg, M. (2001).
\newblock Ordered treemap layouts.
\newblock In {\em IEEE Symposium on Information Visualization, 2001. INFOVIS
  2001.}, pages 73--78. IEEE.

\bibitem[Sugiyama et~al., 1981]{SugiyamaTT81}
Sugiyama, K., Tagawa, S., and Toda, M. (1981).
\newblock Methods for visual understanding of hierarchical system structures.
\newblock {\em {IEEE} Transactions on Systems, Man and Cybernetics},
  11(2):109--125.

\bibitem[Vehlow et~al., 2017]{VehlowBW17}
Vehlow, C., Beck, F., and Weiskopf, D. (2017).
\newblock Visualizing group structures in graphs: A survey.
\newblock In {\em Computer Graphics Forum}, volume~36, pages 201--225. Wiley
  Online Library.

\bibitem[Venn, 1880]{Venn80}
Venn, J. (1880).
\newblock On the diagrammatic and mechanical representation of propositions and
  reasoning.
\newblock {\em Philosophical Magazine Series 5}, 10(59):1--18.

\bibitem[von Hanxleden et~al., 2014]{vonHanxledenDM+14}
von Hanxleden, R., Duderstadt, B., Motika, C., Smyth, S., Mendler, M., Aguado,
  J., Mercer, S., and O'Brien, O. (2014).
\newblock {SCCharts: Sequentially Constructive Statecharts} for safety-critical
  applications.
\newblock In {\em Proc.\ ACM SIGPLAN Conference on Programming Language Design
  and Implementation (PLDI '14)}, pages 372--383, Edinburgh, UK. ACM.

\bibitem[Walshaw, 2001]{Walshaw00}
Walshaw, C. (2001).
\newblock A multilevel algorithm for force-directed graph drawing.
\newblock In {\em Graph Drawing: 8th International Symposium, GD 2000 Colonial
  Williamsburg, VA, USA, September 20--23, 2000 Proceedings 8}, pages 171--182.
  Springer.

\end{thebibliography}

\appendix
\section*{\uppercase{Appendix}}
\label{sec:appendix}
\begin{figure}[b]
	\centering
	\includegraphics[width=\linewidth]{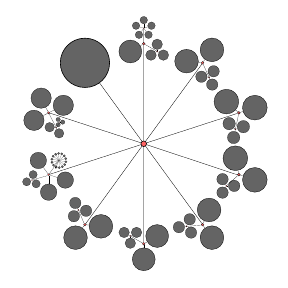}
	\caption{A balloon tree created by applying radial layout in a top-down drawing of a compound graph.}
	\label{fig:balloon-tree}
\end{figure}

\paragraph{Drawing Balloon Trees with Top-Down Layout}
\label{sec:balloon-trees}
In \Cref{sec:related} we stated that top-down layout could be used to create drawings similar to balloon trees. We now show how this can be done. Our input is a tree \(G=(V,E)\) with a root node \(r\). We now transform this tree into a compound graph in the following manner. We create a new compound node \(r'\) and add \(r\) as well as its adjacent nodes as children to \(r'\). We repeat this process recursively for all adjacent node of \(r\). In the case that a node does not have its own unprocessed nodes, i.e., it is a leaf of the tree we do not create a compound node but simply add it as leaf node. In the rendering leaf nodes are drawn as circles, compound nodes are invisible except for a central node drawn in red. Each compound node is laid out using a radial layout algorithm with the red core node acting as the root. Top-down layout scales down the drawings so that they fit inside the compound node parents. We draw hierarchy crossing edges to connect the compound nodes with each other. The result can be seen in \autoref{fig:balloon-tree}.
\end{document}